\documentclass[aps,twocolumn,showpacs,floatfix]{revtex4}

\usepackage{amssymb,amsmath}
\usepackage{graphicx}
\usepackage{subfigure}

\newif\ifpdf
\ifx\pdfoutput\undefined
\pdffalse 
\else
\pdfoutput=1 
\pdftrue
\fi
\ifpdf
\else
\fi

\begin{document}

\title{Automated Chirp Detection with Diffusion Entropy:\\ Application to Infrasound from Sprites}
\author{M. Ignaccolo$^{1,}$\footnote{Correspondent Author. Email: eexmi@bath.ac.uk. \\ Web Page: \\http://staff.bath.ac.uk/eexmi/mywebpage/home/mywebpage.htm}}
\author{T. Farges$^{2}$}
\author{E. Blanc$^{2}$}
\author{M. F\"{u}llekrug$^{1}$}

\affiliation{$^{1}$Department of Electronic and Electrical Engineering, University of Bath, UK}
\affiliation{$^{2}$Commisariat \`{a} l'Energie Atomique, DASE, Bruy\`{e}res le Ch\^{a}tel, France.}
\date{\today}


\begin{abstract}
We study the performance of three different methods to automatically detect a chirp in background noise. $(1)$ 
The standard deviation detector uses the computation of the signal to noise ratio. $(2)$ The spectral 
covariance detector is based on the recognition of the chirp in the spectrogram. $(3)$ The CASSANDRA 
detector uses diffusion entropy analysis to detect periodic patterns in noise. All three detectors are applied 
to an infrasound recording for detecting chirps produced by sprites. The CASSANDRA detector provides the 
best trade off between the false alarm rate and the detection efficiency.
\end{abstract}
\pacs{05.45.Tp,89.20.-a,52.80.Mg}
      \bigskip
\mbox{}

\maketitle

\section{Introduction}\label{intro}

Chirps are periodic signals with an instantaneous frequency changing in time. Chirps are produced 
by a variety of sources: from lightning generated whistlers \cite{whistlers} to the 
acoustic emission of bats \cite{bat1} and whales \cite{whale1}. Recently the presence of chirps in infrasound recording 
has been associated with the occurrence of sprites over 
thunderstorm clouds \cite{thomas}. Several methods of chirp detection have been developed, operating in 
both the time domain (multiple frequency tracker \cite{mft1} and recursive least square algorithm \cite{rls1}) 
and the frequency domain (Page's test \cite{pt1} and 
Hough transform \cite{hough1,hough2}).

In this work  we introduce two new methods for chirp detection. The spectral covariance detector and the 
CASSANDRA detector. The spectral covariance detector operates in the frequency domain and uses the correlation between 
different frequency bins of the spectrogram \cite{mallat} as an indicator of the chirp occurrence. The CASSANDRA 
(\emph{Complex Analysis of Sequences via Scaling AND Randomness Assessment}) detector operates 
in the time domain \cite{cassandra1,cassandra2} and uses diffusion entropy analysis \cite{de1,de2}. 
We study the performance of these two detectors and compare the results with a standard signal to noise ratio 
detector (standard deviation detector).

Sprites \cite{firstsprite} and other recently discovered Transient Luminous Events (TLEs) above thunderstorm clouds, 
such as elves \cite{firstelf}, blue jets 
\cite{firstbj} and gigantic jets \cite{firstgigj} are the subject of intense research 
\cite{natobook}. TLEs connect the lower layer of the 
atmosphere (troposphere: below $10$ km) where the weather activity occurs with the  upper  levels of the 
atmosphere (80-100 km). Knowledge of the sprite occurrence rate is of primary interest 
to address the relevance of these phenomena and their global impact on the atmosphere. 

The outline of this work is the following. In Section \ref{sprite} we briefly discuss the infrasound signature of sprites. 
In Section \ref{detectors} we discuss the details of each detector and its performance in sprite detection. 
In Section \ref{conclusion} we draw our conclusions.

\section{Sprites and their Infrasound Signature}\label{sprite}

Sprites are TLEs with a typical duration from a few milliseconds up to a few hundred milliseconds. They are generated by 
the electric field pulse of a ``parent'' positive cloud-to-ground (+CG) lightning discharge \cite{boccippio}. 
The vertical extension of sprites is $\simeq$45 km, starting from $\simeq$40 km up to $\simeq$85 km, while their 
horizontal extension can range from 20-50 km. Since the first optical observations \cite{firstsprite}, Sprites have been 
observed over thunderstorm clouds in North America 
\cite{optna}, Europe \cite{opteu} and Japan \cite{optjp}. Electromagnetic signatures from sprites have been reported in the 
Extremely-Low Frequency (ELF) range (10Hz-3kHz) \cite{elfsig} and with Earth-ionosphere cavity resonances \cite{srsig}.

The possibility that sprites could generate an infrasound signature was first suggested by Lizska 
\cite{firstinfrahypo}. The first report of sprite signature is by \cite{thomas}. The sprite signatures 
are located in the 1-10 Hz frequency range and in many cases a linear chirp of increasing frequency 
with time is observed. This signature is caused by  the spatial extent of the sprite (from 20 to 50 km) \cite{thomas}, 
its orientation with respect to the infrasound station, and the reflectivity properties of the thermosphere \cite{blanc}. 
Pressure waves generated from different regions of the 
sprite will be reflected at different altitudes in the thermosphere with different absorption and dispersion properties 
before reaching the infrasound station. The net result is that pressure waves coming from the nearest end of the sprite 
will arrive first at the station with a low frequency content. Pressure waves coming from the farthest end of 
the sprite will arrive later at the station with a high frequency content. Sprite signatures which show an impulsive 
feature instead of a chirp, are the result of a small spatial extension or of the alignment with the infrasound 
station (regardless of spatial extent). 

\begin{figure}[h]
\includegraphics[angle=-90,width=\linewidth] {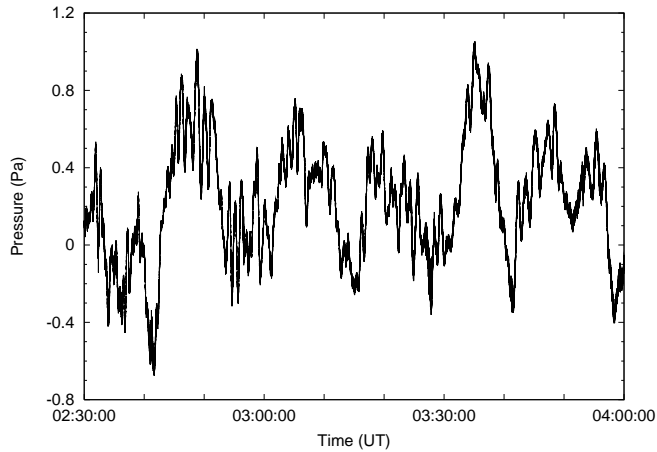}
\caption{The infrasound recording at the Flers station from 2:30 to 4:00 UT on the $21$st of July $2003$. The 
values represent the pressure fluctuations (in Pascal) registered by a micro-barograph.}
\label{figure1}
\end{figure}

The data set used to test the automated chirp detectors is shown in Fig.~\ref{figure1}. It is the original signal recorded with 
a 20 Hz sampling rate at the  infrasound station in Flers (210 km West of Paris, France) from 2:30 UT to 4:00 UT on the 
21st of July 2003. During this hour and half an intense thunderstorm occurred in central France. Optical 
observations reported 28 sprites in the thunderstorm region. From these 28 sprites, 12 signatures 
were detected  in the infrasound recording, $9$ signatures were chirps with an average duration $\bar{T}$$\simeq$12-15 
seconds as identified by visual inspection.

\begin{figure}[h]
\includegraphics[angle=-90,width=\linewidth] {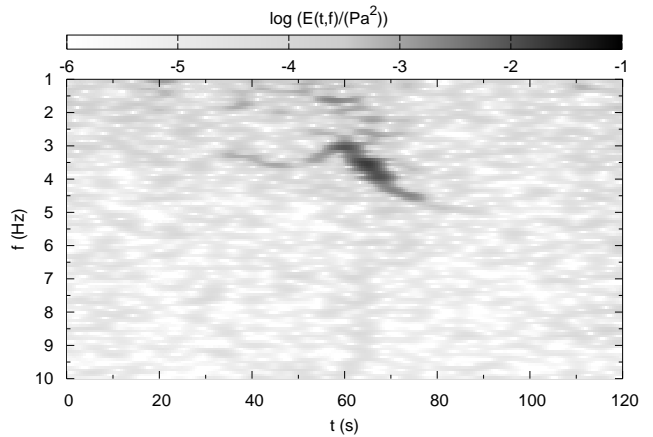}
\caption{A chirp signature of a sprite is identified plotting the energy density (spectrogram) $E(t,f)$ of the signal in 
the time-frequency plane for the 1-10 Hz frequency range. This plot refers to the 2 minute long time interval of 
the original data starting at 2:55:30 UT. The time resolution is 0.5 s and the frequency resolution is 1/256 Hz.
 This signature exhibits an extremely good signal to noise ratio.}
\label{figure2}
\end{figure}

The sprite signatures are not visible in Fig.~\ref{figure1} because their intensity is very small compared to those of 
the slow pressure  ($f$$\lesssim$1 Hz) fluctuations caused by the wind. But the spectrogram in the range $1-10$ Hz as 
in Fig.~\ref{figure2} shows the chirp signature of a sprite. Therefore before applying any detection method, we high pass filter 
the infrasound recording in order to eliminate the very slow wind fluctuations ($f$$\lesssim$1 Hz). 

\section{Automated detectors and their performance}\label{detectors}

The performance of an automated detector operating with a threshold $\tau$ is measured by the Detection Efficiency $DE(\tau)$ 
and by the False Alarm rate $FA(\tau)$ . The number of sprites occurring $N_{O}$ can be obtained from the number of 
sprites detected $N_{D}(\tau)$ by
\begin{equation}\label{halfprose}
N_{O} = N_{D}(\tau) \;\; \frac{1-FA(\tau)}{DE(\tau)}.
\end{equation}
The ideal detector has a threshold $\tau$ with no false alarm rate ($FA(\tau)=0$) 
and perfect detection efficiency ($DE(\tau)=1$). This implies $N_{O}=N_{D}(\tau)$. The optimal threshold $\tau$ is the best 
compromise between detection efficiency and false alarm rate. 
A threshold with zero false alarm rate and a small detection efficiency is not desirable because it means that only 
rare events are detected (perfect chirp signals unaffected by noise). In this case the number of detected occurrences are 
affected by noise. The same is true for a perfect detection efficiency but a large false alarm rate where many detections 
result from signatures other than chirps.

Therefore we investigate the properties of the detection efficiency (DE) and the false alarm rate (FA) as a 
function of the threshold $\tau$ for each detector. 

\subsection{The standard deviation detector}

\begin{figure}[h]
\includegraphics[angle=-90,width=\linewidth] {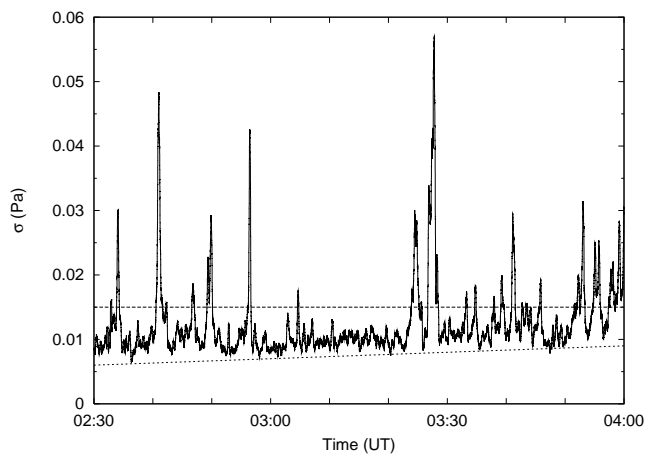}
\caption{The variation of the standard deviation $\sigma$ of the high pass filtered signal in a 12.8 second long interval. 
The horizontal dashed line indicates a threshold of $0.015$ Pa. The diagonal dotted line at the bottom shows the 
increase in intensity of the pressure fluctuations as sun rise approaches.}
\label{figure3}
\end{figure}

The standard deviation detector detects  a chirp when the signal intensity exceeds a given threshold. We calculate 
the signal intensity changes in the high pass filtered infrasound recording, 
by moving a window of length $L_{s}$ through the data and evaluating at each time the standard deviation of the data inside 
the window. We choose $L_{s}= 256$ data points. This corresponds to a $12.8$ seconds long time interval: a time interval in 
the expected range of the average duration ($\bar{T}$$\simeq$12-15 s) of a sprite chirp signature. 
The variation of the standard deviation is shown in Fig.~\ref{figure3}. 
The horizontal line indicates the threshold $\tau = 0.015$ Pa., while the bottom diagonal line indicates the long term 
trend of the standard deviation. 
The intensity follows the day-night cycle of temperature, with a maximum around noon and a minimum around midnight. 
In Fig.~\ref{figure3} the intensity slowly increases as the sun rise approaches at 4:00 UT.

A detector based on the standard deviation should continuously scale the threshold to take in account the 
effect of change of intensity in time or particularly windy conditions. Here we use
 the standard deviation detector for comparison with the spectral 
covariance detector and the CASSANDRA detector. Therefore we use a constant threshold for the entire duration (2 hours) 
of the infrasound recording.

\begin{figure}[h]
\includegraphics[angle=-90,width=\linewidth] {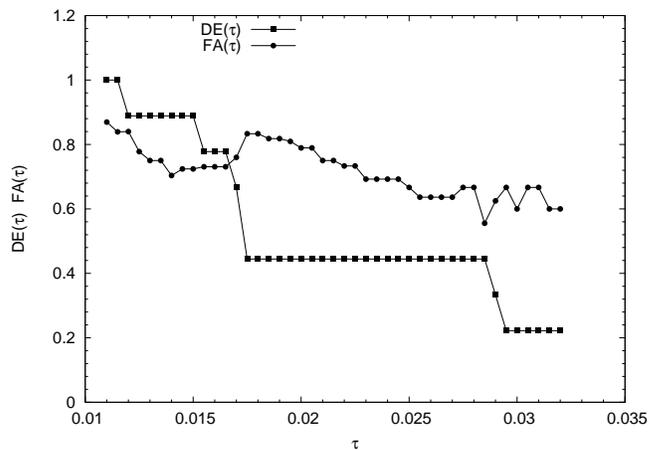}
\caption{Standard deviation detector: the Detection Efficiency (squares) and the False Alarm rate (circles) as a function of 
the threshold $\tau$.}
\label{figure4}
\end{figure}

Fig.~\ref{figure4} shows the detection effieciency and the false alarm rate for different values of the threshold $\tau$. It is evident 
that the standard deviation detector has no good compromise between false alarm rate and detection efficiency. For values of  
$\tau \leq 0.017$ we have a detection efficiency (DE $\geq$ 0.6 or 60$\%$) but a false alarm rate (FA $\geq$ 0.75 or 75$\%$). 
Raising the threshold lowers the false alarm rate to about 0.6 (60$\%$) but the detection efficiency decreases to less 
than 0.3 (30$\%$). These results for the standard deviation detector are not surprising: every chirp signature implies an 
increase in the signal to noise ratio but the  inverse conjecture is not true. 


\subsection{The spectral covariance detector}

The spectral covariance detector rests on the application of the spectral covariance of the spectrogram to detect
chirp signatures of increasing frequency (Fig. \ref{figure2}). In this section, we introduce the spectral covariance and 
discuss the performance of the spectral covariance detector for the infrasound chirp signatures of sprites.  

\subsubsection{Spectral covariance}

The Spectrogram of a signal is its energy density $E(t,f)$ in the time-frequency plane. 
A linear chirp of increasing frequency will appear as a diagonal line in the spectrogram (Fig.~\ref{figure2}). 
The inclination of the line with respect the horizontal axis is proportional to the rate $v$ at which the frequency 
of the chirp changes in time. The spectral covariance uses the covariance between the energy density $E(t,f)$ relative 
to two different frequencies to detect the presence of a diagonal line in the spectrogram. 
The covariance of time delay $\delta$  relative to the frequencies $f$ and $f+f_{s}$ is 
\begin{equation}\label{cov}
C_{f,f_{s}} (\delta) = <E(t,f)E(t+\delta, f+f_{s})>_{t},
\end{equation}  
where $<...>_{t}$ denotes the time average and $f_{s}$$>$$0$ is the frequency shift.

For a white noise signal, the computed spectrogram $E(t,f)$ is a randomly fluctuating function in both the arguments $t$ and $f$.  
Thus, every delay $\delta$ has the same probability of maximizing the covariance $C_{f,f_{s}} (\delta)$. 
The same holds true if a periodic component of fixed frequency $f_{o}$ is superimposed on the noise. 
In this case, if both $f$ and $f+f_{s}$ are different from $f_{o}$ then $E(t,f)$ and 
$E(t+\delta,f+f_{s})$ are randomly fluctuating functions, if $f$=$f_{o}$ then $E(t,f_{o})$ is constant but $E(t+\delta,f+f_{s})$ is a 
randomly fluctuating function and vice-versa. In the case of a linear chirp with initial frequency $f_{i}$ and final frequency 
$f_{f}>f_{i}$ 
the covariance $C_{f,f_{s}} (\delta)$ will have a maximum at $\delta_{max} \neq 0$ whenever the frequencies $f$ 
and $f+f_{s}$ are in the range $\left[f_{i},f_{f}\right]$ or, equivalently, 
\begin{equation}\label{upperboundfs}
f_{s} < f_{f} - f_{i}.
\end{equation}
The value $\delta_{max}$ is inversely proportional to the 
rate $v$ at which the frequency of the chirp is changing:
\begin{equation}\label{speed}
\delta_{max}=\frac{f_{s}}{v}.
\end{equation}
Finally, in the case of an impulsive signature (a straight vertical line) in the spectrogram like the one of
the thunder produced by a lightning, we expect $\delta_{max}$$=$$0$ ($v$$\rightarrow +\infty$). 

\begin{figure}[h]
\includegraphics[angle=-90,width=\linewidth] {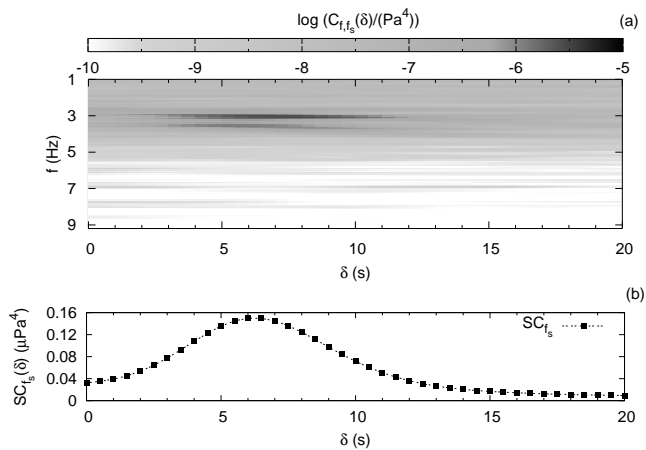}
\caption{(a) The covariance $C_{f,f_{s}} (\delta)$ as a function of the delay $\delta$ (in seconds) for the spectrogram of Fig. 2. 
Here $f_{s}$=10/256 Hz ($\simeq$0.8 Hz). (b) The spectral covariance $SC_{f_{s}}$ as a function of the delay $\delta$ (in seconds).}
\label{figure5}
\end{figure}

To use the covariance of Eq.~(\ref{cov}) for detecting linear chirps with a variable frequency range we need to 
eliminate its dependence on a particular frequency $f$ without loosing the useful properties in chirp detection. 
Thus we average the covariance over all frequencies and define the spectral covariance of 
delay $\delta$ and frequency shift $f_{s}$ as  
\begin{equation}\label{spcov}
SC_{f_{s}} (\delta) = <C_{f,f_{s}} (\delta)>_{f},
\end{equation}
where $<...>_{f}$ denotes the frequency average. The delay $\delta_{max}$ for which the spectral covariance of Eq.~(\ref{spcov}) 
holds its maximum value, has exactly the same properties of the delay $\delta_{max}$ relative to the covariance of Eq.~(\ref{cov}). 
In Fig.~\ref{figure5} we plot the covariance $C_{f,f_{s}}$ and the spectral covariance $SC_{f_{s}}$ as a function of the delay 
$\delta$ for the 2 minute long spectrogram of Fig.~\ref{figure2}. The chirp of Fig.~\ref{figure2} produces large values of the 
covariance for frequencies $f$ between 3 and 4 Hz and delays $\delta$ between 5 and 10 seconds. The spectral covariance has its 
maximum for $\delta_{max}$$\simeq$6 seconds. This value of $\delta_{max}$ corresponds (Eq.~\ref{speed}) to a rate of frequency 
change $v$$\simeq$0.13 Hz/s and an inclination of the chirp signature in the spectrogram of $\simeq$58$^{o}$ (Fig.~\ref{figure2}).  

\subsubsection{Chirp detection}

The spectral covariance detector operates as follows. An interval of the spectrogram of duration $I$ centered in the location $t$ 
is examined and the delay $\delta_{max}(t)$ evaluated. Then the interval is shifted and centered 
to the new location $t+u$ and the correspondent $\delta_{max}(t+u)$ is evaluated. Consecutive intervals containing a 
linear chirp of increasing frequency will result in consecutive equal values of the delay 
$\delta_{max}(t)\neq0$. For a linear chirp of time duration $\bar{T}$, the number $n_{CEV}$ of consecutive equal values of the delay 
$\delta_{max}$ is in the range
\begin{equation}\label{ncev}
\frac{\bar{T}}{u} \leq n_{CEV} \lesssim \frac{I}{u}. 
\end{equation}
In real cases a chirp signature in the spectrogram will produce similar consecutive values of the delay $\delta_{max}(t)$. 
The automated detection algorithm measures the dispersion of $n_{CEV}$ consecutive values of $\delta_{max}(t)$ around their mean. 
A detection will be reported if the relative dispersion (the ratio between the standard deviation and the average) 
is below a given threshold $\tau$.       

There are some numerical limitations in the evaluation of the spectral covariance $SC_{f_{s}}$ which must be considered. 
 (1) Numerically, one evaluates the energy contained in a box of dimension $\Delta t$$\times$$\Delta f$ 
centered in the location $(t,f)$ of the time-frequency plane \cite{mallat}. The numerical energy density $E_{NUM}(t,f)$ is obtained dividing 
the energy by the dimension of the box. The locations $(t,f)$ for which the numerical energy density $E_{NUM}(t,f)$ is computed belong to a grid 
of steps $\delta t$ in the time domain and $\delta f$ in the frequency domain. The intervals $\delta t$ and $\delta f$ are the time and frequency 
resolution of the spectrogram, while the intervals $\Delta t$ and $\Delta f$ are the time and frequency localization of the spectrogram. 
Typically, $\Delta t$$>$$\delta t$ and $\Delta f$$>$$\delta f$. These numerical limitations impose a lower bound on the frequency shift $f_{s}$ 
\begin{equation}\label{lowerboundfs}
f_{s} > \Delta f.
\end{equation}
If the condition of Eq.~(\ref{lowerboundfs}) is not satisfied, the numerical energy densities $E_{NUM}(t,f)$ and $E_{NUM}(t,f+f_{s})$ used 
in the evaluation of the spectral covariance (Eq.~\ref{spcov}) refer to two overlapping intervals of frequencies. In this 
case Eq.~(\ref{speed}) may not be satisfied. (2) The duration $I$ imposes a limitation on the values
of delays $\delta$ for which a statistically meaningful numerical evaluation of the spectral covariance, is possible. 
If $\delta_{R} \lesssim I$ is the maximum delay used in evaluating the spectral covariance, only linear chirps 
with a rate of frequency change of
\begin{equation}\label{chirpdetectable}
v > \frac{f_{s}}{\delta_{R}}  
\end{equation}
can be detected.

\subsubsection{Application to infrasound from sprites}

A visual inspection of the chirp signatures of the $9$ sprites present in the infrasound recording examined, shows that the
difference between the final and the initial frequency of the chirps is $\approx$2-2.5 Hz. We evaluate the spectrogram with a frequency 
resolution of 1/256 Hz and we set the frequency shift $f_{s}$ to 10/256 Hz (0.78125 Hz). This value of the frequency shift satisfies 
the inequalities of Eqs.~(\ref{upperboundfs}) and (\ref{lowerboundfs}). We select $I$=30 s and $\delta_{R}$=20 s. With this value 
of $\delta_{R}$ only chirps with a rate of frequency change $v$ greater than 0.039 Hz/s can be detected. These chirps will produce 
signatures of inclination greater than $27^{o}$ in a 2 minutes long display of the spectrogram. 
The shift $u$ of two consecutive intervals of duration $I$ is chosen to be equal to 3.2 seconds and consequently 
(Eq.~\ref{ncev}) we set $n_{CEV}$=7. 
Finally we want to exclude the possibility to detect impulsive signature from lightning. This signatures 
in theory should produce a $\delta_{max}$=0 s, corresponding to vertical lines (inclination of $90^{o}$) in the spectrogram. 
In practice, however, it is better to exclude $\delta_{max}$ too close to zero. We consider only those 
delays $\delta_{max}$$>$1.8 s corresponding to an inclination of $\simeq$$80^{o}$ in a 2 minute long display of the spectrogram.  

\begin{figure}[h]
\includegraphics[angle=-90,width=\linewidth] {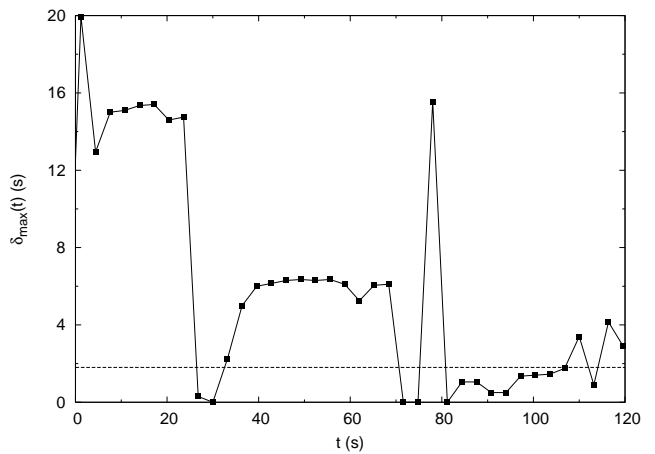}
\caption{The delay $\delta_{max}$ (in seconds) for a interval of duration $I$ of the spectrogram centered 
at the location $t$ (in seconds). This plot refers to the same 2 minute long interval of the original data used for Fig.~2. 
The dashed line at the bottom indicates the delays $\delta_{max}$$=$1.8 s expected for a signature with a 
inclination of $\simeq$$80^{o}$ in a 2 minutes long spectrogram.}
\label{figure6}
\end{figure}

\begin{figure}[h]
\includegraphics[angle=-90,width=\linewidth] {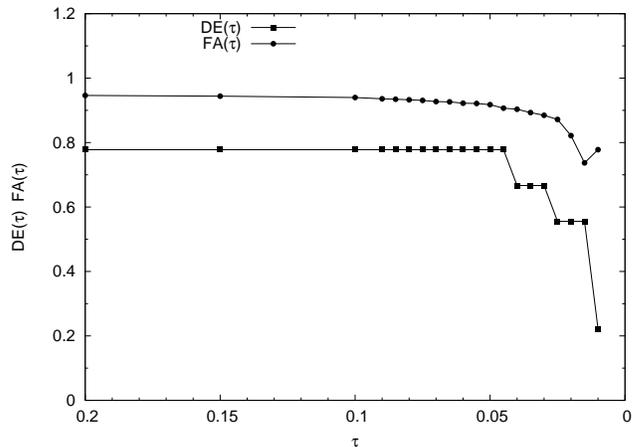}
\caption{Spectral covariance detector: the Detection Efficiency (squares) and the  False Alarm rate (circles) as a function of 
the threshold $\tau$.}
\label{figure7}
\end{figure}

In Fig.~\ref{figure6}, we plot the sequence of delay $\delta_{max}(t)$ for the same 2 minute long time interval of the infrasound 
recording  used for Fig.~\ref{figure2}. The presence of the chirp coincide with consecutive almost equal values of the 
delay $\delta_{max}(t)$ (around $t$=60 s). Before and after the chirp signature a small number of consecutive similar values  
of $\delta_{max}(t)$ (around $t$=20 s and $t$=100 s) and some isolated fluctuating values. 
In Fig.~\ref{figure7}, we plot the detection efficiency and the false alarm rate for different values of the threshold $\tau$. 
As for the standard deviation detector there is no good compromise between the false alarm rate and the detection efficiency. 
For values of $\tau>0.1$ the spectral covariance detector has a detection efficiency of almost 80\%, but a false alarm rate superior to 
90\%. Lowering the threshold results in a false alarm rate slightly below 80\%, but in a drop of the detection efficiency from 
$\simeq$80\% to $\simeq$20\%. This behavior of the detection efficiency and of the false alarm rate has two causes. (1) ``Spurious'' 
signatures (not from sprite) produce a sequence of values of $\delta_{max}$ with a small ($\leq$0.05 or 5\%) relative dispersion. 
(2) Sprites signatures may produce sequence of delay $\delta_{max}$ with an large ($>$0.1 or 10\%) relative dispersion.     

\subsection{The CASSANDRA detector}

The detector described in this section derives its name from the CASSANDRA analysis \cite{cassandra1,cassandra2}: an application of 
diffusion entropy analysis \cite{de1,de2} to non stationary time series \cite{myneurons}. 
In the following we briefly discuss the details of the diffusion entropy analysis and the changes to the 
original \cite{cassandra1,cassandra2} formulation of the CASSANDRA analysis necessary for the detection of chirps. 
We then show the results of the application of the CASSANDRA detector to the infrasound chirp signature of sprites.

\subsubsection{Diffusion entropy analysis}
The application of the diffusion entropy analysis to a time series is made of two steps. 
Step 1:  use the time series to create a diffusion process. Step 2: monitor, as diffusion takes place, the 
entropy of the probability density function (pdf) describing the diffusion process. 
The behavior of the  entropy is indicative of the statistical properties of the time series analyzed. 

The first step of the diffusion entropy analysis is computing all the 
possible sums of any $n$ consecutive terms of the time series 
$\left\{ \xi_{j}\right\}$ of length $L$, namely:

\begin{eqnarray}
&x_{k,n}= \sum\limits_{j=k}^{k+n-1} \xi_{j} \label{aggregation} \\
&k=1,...,L-n+1.\;\;\;\text{ and }\;\;\;n=1,...,L. \nonumber  
\end{eqnarray}
This procedure describes a diffusion process if we consider the sequence $\xi_{j}$ as the sequence 
of fluctuations of a diffusing trajectory and $n$ as the time for which the diffusion process has taken place. 
Consequently, each of the  $L$$-n$$+1$ values $x_{k,n}$ can be thought as the position of a diffusion trajectory 
after a time $n$ starting from the location $0$ at $n$$=$$0$. The second step is computing the pdf $\rho(x,n)$ of finding 
a trajectory in the location $x$ after a time $n$ and its diffusion entropy
\begin{equation}\label{diffusionentropyth}
S(n) = - \int \rho(x,n) \ln \rho(x,n) dx.
\end{equation}
The numerical evaluation of $S(n)$ is done by dividing at each time $n$ the diffusion space in cells of 
equal size $\Delta(n)$ centered around the location $x_{j}$. The size $\Delta(n)$ must be small enough for the 
pdf $\rho(x,n)$ to be  constant inside the cell. In this case, 
\begin{equation}\label{diffusionentropynum}
S(n) \simeq - \sum_{j} p(x_{j},n) \ln p(x_{j},n)+\ln \Delta(n) \\
\end{equation}
with
\begin{equation}\label{prho}
p(x_{j},n) = \rho(x_{j},n) \Delta(n) \nonumber.
\end{equation}
The choice of having a temporal dependence on the cell size $\Delta$ is due to the necessity of satisfying the condition
$\rho(x_{j},n)$ constant inside a cell and the probability $p(x_{j},n)$ being large enough for a meaningful 
statistical evaluation with $L$$-n$$+1$ trajectories at time $n$ (Eq.~\ref{aggregation}). A choice of a 
small fixed size cell will satisfy the former condition but not the latter when $n$ increases and the diffusion 
trajectories explore larger intervals of the diffusion space. 

A time series $\left\{\xi_{j}\right\}$ of random uncorrelated numbers drawn from a distribution of finite variance generates 
a diffusion process that rapidly becomes Brownian \footnote{This is a consequence of the Central Limit Theorem.}. As a consequence, the diffusion 
entropy $S(n)$ will rapidly approach a regime of linear increase on a logarithmic time scale with 
slope 0.5 \footnote{For a Brownian diffusion the pdf $\rho(x,n)$ can be written as $\frac{1}{t^{\delta}} F(\frac{x}{t^{\delta}})$ 
with $\delta=0.5$ and the function $F(y)$ a Gaussian. In this case, a straightforward calculation shows that $S(n)=A+\delta \ln n$, 
with $A$ a constant and $\delta=0.5$.}.
 
\begin{figure}[h]
\includegraphics[angle=-90,width=\linewidth] {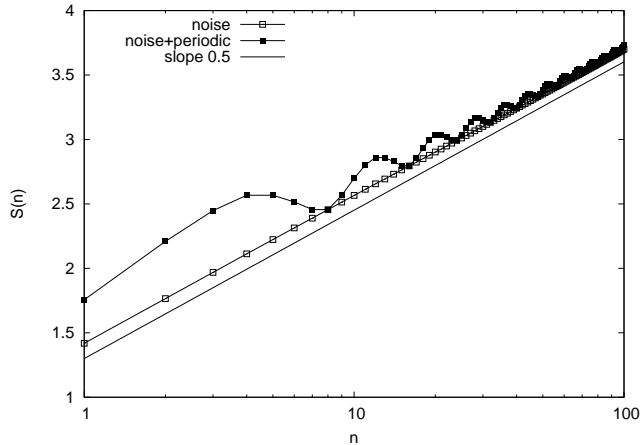}
\caption{The diffusion entropy $S(n)$ as function of time $n$ of gaussian noise (line and white squares) and of gaussian noise plus 
a periodic component (line and black squares). The periodic component has a periodicity of  $M$ of 8 data points. 
The full line represents the increase of $S(n)$ with a slope of $0.5$ on a logarithmic time scale.}
\label{figure8}
\end{figure}

The addition to the random time series $\left\{\xi_{j}\right\}$ of a periodic component 
$\left\{\xi^{M}_{j}\right\}$ with a periodicity of $M$ data points, has the effect of ``bending'' the diffusion 
entropy: the periodic component increases the value of $S(n)$ at times $n$ that are not multiples of $M$ 
and has no effect on  the value of $S(n)$ at times $n$ multiples of $M$. As a consequence, the times $n$ multiple of 
the period $M$ are now points of local minima for the diffusion entropy $S(n)$. The ``bending'' effect is shown 
clearly in Fig. \ref{figure8}. When a periodic component $\left\{\xi^{M}_{j}\right\}$ is added to a random noise fluctuation 
$\left\{\xi_{j}\right\}$ we can write the sum $x_{k,n}$ of Eq.~(\ref{aggregation}) as
\begin{align}
x_{k,n}= & \sum\limits_{j=k}^{k+n-1}\left( \xi_{j}+\xi^{M}_{j} \right) \nonumber \\
= & \sum\limits_{j=k}^{k+n-1} \xi_{j}+\sum\limits_{j=k}^{k+n-1} \xi^{M}_{j}=x_{k,n}^{noise}+x_{k,n}^{periodic} \label{aggregationperiodic} \\
&k=1,...,L-n+1.\;\;\;\text{ and }\;\;\;n=1,...,L. \nonumber  
\end{align}
The addition of the periodic component $\left\{\xi^{M}_{j}\right\}$ has the effect of shifting the position $x_{k,n}^{noise}$ at time $n$ of
the $k$-th diffusion trajectory by the quantity $x_{k,n}^{periodic}$. If $n$ is not a multiple of the period $M$, the shift 
$x_{k,n}^{periodic}$ depends on the index $k$. Therefore different diffusion trajectories are shifted by a different quantity. This 
results in a bigger ``spreading'' of the diffusion trajectories and therefore in a larger value of the diffusion entropy $S(n)$. When $n$ 
is a multiple of the period $M$, the shift $x_{k,n}^{periodic}$ is independent from the index $k$. In this case all the diffusion 
trajectories are shifted by the same quantity and the diffusion entropy $S(n)$ does not change. 

Moreover, Fig.~\ref{figure8} shows how the ``bending'' of the diffusion entropy $S(n)$ becomes smaller and smaller 
as the time $n$ increases. The standard deviation of the sum $x_{k,n}^{noise}$ (Eq.~\ref{aggregationperiodic}) of $n$ consecutive 
terms of the random fluctuations $\left\{\xi_{j}\right\}$ increases as $n$ increases. The standard deviation of the sums 
$x_{k,n}^{peridodic}$ (Eq.~\ref{aggregationperiodic}) of $n$ 
consecutive terms of the periodic component $\left\{\xi^{M}_{j}\right\}$ is limited and it is periodic of period $M$: it vanishes 
whenever $n$ is a multiple of the period $M$, it increases, reaches a maximum and then decreases in between two consecutive periods.
Therefore the contribution to the diffusion entropy $S(n)$ of the periodic component becomes 
smaller and smaller compared to that of the random values $\left\{\xi_{j}\right\}$ and so does the ``bending'' effect. 

\subsubsection{CASSANDRA analysis}

The CASSANDRA analysis is the application of the diffusion entropy analysis to smaller 
intervals of a time series such that the differences between 
the results in each interval reflect the statistical changes occurring in the time series itself. To compare the differences between 
the diffusion entropy of different intervals of a time series, the authors of \cite{cassandra1,cassandra2,myneurons} use the following 
quantity
\begin{equation}\label{origcassandra}
C_{L_{s}}(t) =  \frac{1}{N} \sum_{n=1}^{N} \left[ S_{t}(n) - \left(S_{t}(1)+0.5 \ln(n) \right) \right].
\end{equation}
In Eq.(\ref{origcassandra}), $t$ is the position in the time series where the small interval of length $L_{s}$ is centered, 
$S_{t}$ is the diffusion entropy relative to this interval of the time series and $N$ is the maximum time for which the 
evaluation of $S_{t}(n)$ is statistically meaningful. $C_{L_{s}}(t)$ is the difference between the ``local'' diffusion entropy 
$S_{t}(n)$ and one that, starting from the same value at $n=1$, increase with a slope of 0.5 on a logarithmic time scale. 
Therefore, $C_{L_{s}}(t)$ is an indication of how different the diffusion process generated by the data in the small interval 
centered at $t$ is from Brownian diffusion. The quantity $C_{L_{s}}(t)$ of Eq.~(\ref{origcassandra}) is useful in detecting 
increases of the diffusion entropy $S_{t}(n)$ with an average slope smaller or larger than 0.5 as a result of the local 
correlation properties. But in our case, we want to detect chirps in background noise. Intervals containing a chirp 
will result in a ``bended'' diffusion entropy  with the times $n$ of the local minima of $S_{t}(n)$ depending on the 
chirps instantaneous frequency. For this reason we evaluate, instead of the quantity defined in Eq.~(\ref{origcassandra}), the 
cumulative slope change $CSC_{L_{s}}(t)$ and the time $\tilde{n}_{L_{s}}(t)$ of the first local minima of $S_{t}(n)$. 

The cumulative slope change for a interval 
of length $L_{s}$ centered at the position $t$ of the time series is defined as   
\begin{equation}\label{csc}
CSC_{L_{s}}(t) = \frac{1}{N-2} \sum_{n=0}^{N-2}\frac{\vert \theta_{t}(n+1) - \theta_{t}(n)\vert}{\ln(n+2)-\ln(n)},
\end{equation}
where
\begin{equation}\label{slope}
\theta_{t}(n)=\frac{S_{t}(n+1)-S_{t}(n)}{\ln(n+1)-\ln(n)}. 
\end{equation}
$S_{t}(n)$ and $N$ of Eq.~(\ref{csc}) are the same quantities as in Eq.~(\ref{origcassandra}), while $\theta_{t}(n)$ 
of Eq.~(\ref{slope}) is the slope of the line connecting two consecutive values of the diffusion entropy $S_{t}(n)$ when 
plotted on a logarithmic time scale. 
The cumulative slope change of Eq.~(\ref{csc}) is the weighted sum of the absolute value of the difference between two consecutive slopes. 
The weights are the intervals of time (on a logarithmic time scale) during which the slope difference is evaluated. 
Slope changes happening at later times are weighted more. This compensates the fact that the ``bending'' 
of the diffusion entropy becomes smaller as $n$ increases. 
The cumulative slope change is able to distinguish intervals of the time series where the diffusion entropy $S_{t}(n)$ is 
``bended'' from those where it is not. The time of the first local minima $\tilde{n}_{L_{s}}(t)$ is the first time $n$ for which the conditions 
$\theta_{t}(n-1)<0$ and $\theta_{t}(n)>0$ are satisfied. If these condition are not met for any $n<N$ then we set 
$\tilde{n}_{L_{s}}(t)=N$. Therefore for an interval of length $L_{s}$ centered at the position $t$ which contains  
only noise $\tilde{n}_{L_{s}}(t)=N$ \footnote{Intervals of a time series including only noise can produce a diffusion entropy 
$S(n)$ with a first local minima $\tilde{n}_{L_{s}}<N$ particularly when $L_{s}$ is so small that the statistics necessary to evaluate 
$S(n)$ is not optimal. Therefore we register only the first times of local minima which are statistically relevant. The first times of 
local minima of $S(n)$ are statistically relevant if $\frac{\theta(n)-\theta(n-1)}{\theta(n)+\theta(n-1)}> \alpha$. In a run with one 
million different sequences of random gaussian noise of length $L_{s}=256$  only 0.1\% where statistically relevant if $\alpha=0.3$.}. 
For a sequence of intervals with a periodic component of fixed period $M<N$, $\tilde{n}_{L_{s}}(t)=M$. Finally for a sequence of intervals 
containing a chirp signature  $\tilde{n}_{L_{s}}(t)$ will change accordingly to the local chirp frequency.
\subsubsection{Chirp detection}

For the purpose of chirps detection we need to use the cumulative slope change $CSC_{L_{S}}(t)$ together with the time of first local minima 
$\tilde{n}_{L_{s}}(t)$. . We define the chirp cumulative slope change $cCSC_{L_{s}}(t)$ as
\begin{equation}\label{chirpcsc}
cCSC_{L_{S}}(t)= \begin{cases}
0 \: \text{if} \begin{cases}
	\tilde{n}_{L_{S}}(t)\!=\!N \\
	\tilde{n}_{L_{S}}(t)\!=\!M\!<\!N \quad t\in J 
	\end{cases} \tag{16}\\
\\
CSC_{L_{s}}(t) \;\; \text{otherwise}
\end{cases}
\end{equation}
The set $J$ is the set of the all the locations $t$ for which consecutive constant values of $\tilde{n}_{L_{S}}(t)=M<N$  are encountered.
The chirp cumulative slope change of Eq.~(\ref{chirpcsc}) vanishes for intervals containing only noise or a periodic component of fixed 
period $M$.
The CASSANDRA detector detects a chirp when the chirp cumulative slope change $cCSC_{L_{s}}(t)$ exceeds a given threshold $\tau$.

\subsubsection{Application to infrasound from sprite}

As for the standard deviation detector we choose $L_{s}=256$ corresponding to a 12.8 seconds long time interval. 
Moreover, we dichotomize the signal such that every data point above the average is $+1$ and every data point 
below the average is $-1$. This drastic procedure has three advantages. (1) The night-day cycle of the intensity in the infrasound 
recording (Fig.~\ref{figure3}) is eliminated by the dichotomization. Thus, a unique threshold independent from the time of 
day can be set for the purpose of automated detection. (2) The problem of choosing an appropriate value of $\Delta(n)$ for 
different times $n$ in Eq.~(\ref{diffusionentropynum}) is simplified: a unitary cell will be used for all the values of $n$ explored. 
(3) The signal to noise ratio is ``preserved'' by the dichotomization: a very intense sprite 
signature will result in an almost perfect periodic pattern with decreasing frequency, while a weak signature 
will be difficult to recognize because the noise will randomly affect the pattern. 

\begin{figure}[h]
\includegraphics[angle=-90,width=\linewidth] {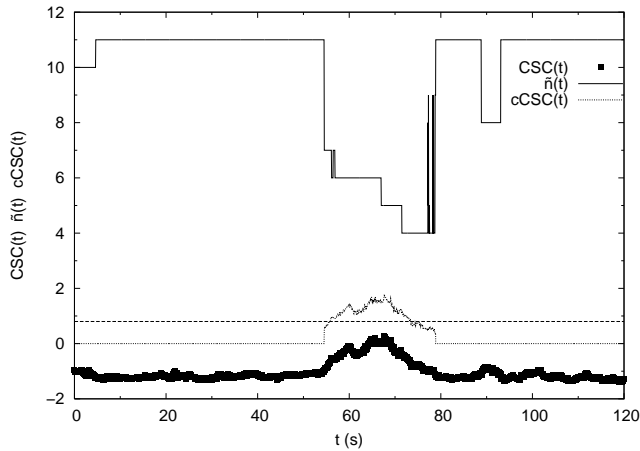}
\caption{The cumulative slope change $CSC_{L_{S}}$ (squares), the time $\tilde{n}$ (full line) of the first local minima of the 
diffusion entropy $S^{t}(n)$  and the chirp cumulative slope change $cCSC_{L_{S}}$ (dotted line) as a function of the position $t$ (in seconds). 
The plot refer to the same 2 minute long interval of the original data used for Fig. 2. The cumulative slope change has been translated down 
for clarity. The horizontal line at the bottom indicates a threshold of 0.8.}
\label{figure9}
\end{figure}

\begin{figure}[h]
\includegraphics[angle=-90,width=\linewidth] {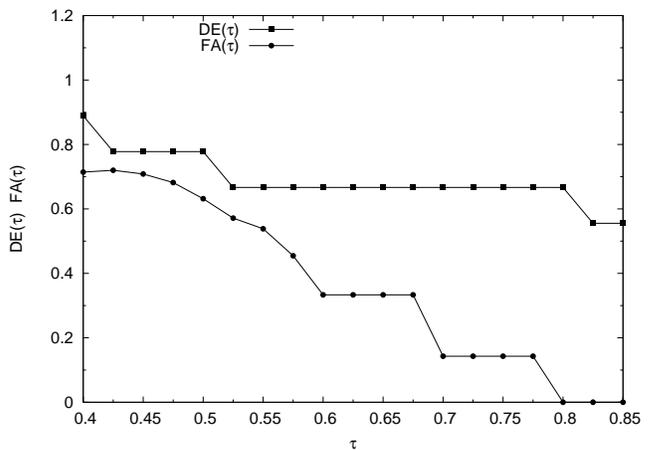}
\caption{CASSANDRA detector: the Detection Efficiency (squares) and the False Alarm rate (circles) as a function of 
the threshold $\tau$.}
\label{figure10}
\end{figure}

In Fig.~\ref{figure9} we plot the chirp cumulative slope change $cCSC_{L_{s}}(t)$, the cumulative slope change $CSC(t)$ (for clarity the 
cumulative slope change has been moved down with respect the  chirp cumulative slope change) and the time of 
the first local minima $\tilde{n}(t)$ for the same insert of 2 minutes used for Fig.~\ref{figure2}. 
We clearly see that the chirp signature produces a big value of the cumulative slope change and that the time 
of first local minima $\tilde{n}_{L_{s}}$ detects its change in frequency passing with continuity from 7 to 4. 
The chirp cumulative slope change of Eq.~(\ref{chirpcsc}) reduces the possibility of a false alarm annulling the cumulative slope 
change in the case of noise or periodic component with fixed periodicity.,  

Finally, in Fig.~\ref{figure10} we plot the the detection efficiency and the false alarm rate for different values of the threshold $\tau$. 
The CASSANDRA detector has a good compromise between the false alarm rate and the detection efficiency. For a threshold value of 0.8 the 
false alarm rate is null and the detection efficiency is about 66\%. 
\section{Conclusion}\label{conclusion}
\begin{figure}[h]
\includegraphics[angle=-90,width=\linewidth] {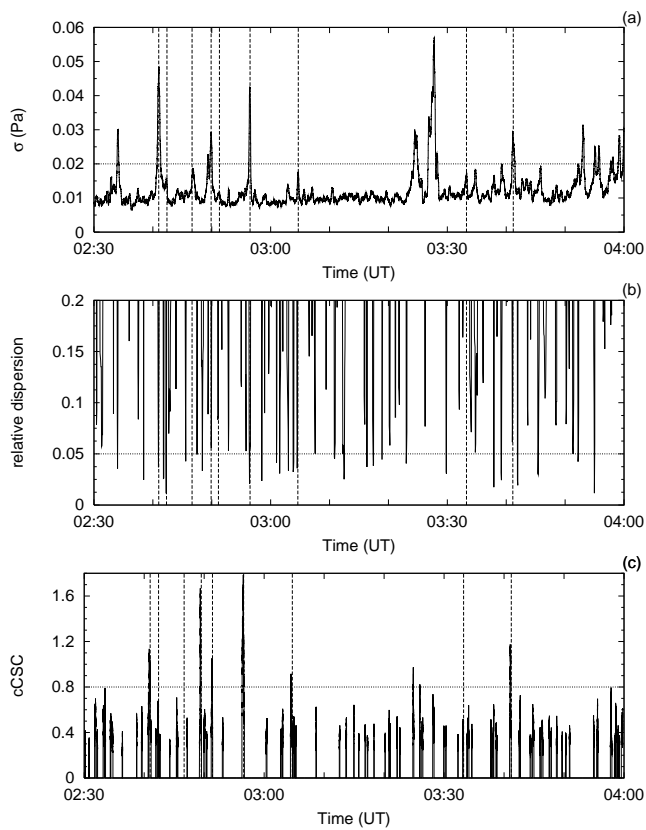}
\caption{Comparison between the three different detectors used. (a) The standard deviation detector. (b) The spectral covariance detector. (c) 
The CASSANDRA detector. In all cases the the dashed vertical line indicate the time of the sprite signatures, while the horizontal dotted lines 
indicates a threshold.}
\label{figure11}
\end{figure}
The plots of the detection efficiency and false alarm rate as a function of the threshold $\tau$ (Figs.~\ref{figure4}, \ref{figure7} and 
\ref{figure10}) indicate
that the CASSANDRA detector is the one with the best trade off between detection efficiency and false alarm rate. This is confirmed by the plot 
of Fig.~\ref{figure11}. We see how raising the threshold $\tau$ in the case of the standard deviation detector (panel (a)) does not lower the 
false alarm rate. For the spectral covariance detector we would expect to get a better false alarm rate lowering the threshold, but this is not 
the case (panel (b)). Finally we see how raising the threshold in the CASSANDRA detector improves the false alarm rate without compromising the 
detection efficiency (panel (c)).

\end{document}